\documentclass[12pt]{article}

\usepackage{amsmath}
\usepackage{latexsym}
\usepackage{amssymb}

\newtheorem{thm}{Theorem}

\newtheorem{Defn}[thm]{Definition}
\newtheorem{Remark}[thm]{Remark}
\newtheorem{Note}[thm]{Note}

\newtheorem{Example}[thm]{Example}
\newtheorem{Examples}[thm]{Examples}
\newtheorem{Problems}[thm]{Problems}

\newtheorem{Problem}[thm]{Problem}
\newtheorem{Notation}[thm]{Notation}
\newtheorem{Number}[thm]{\!\!}

                  {\nopagebreak\hspace*{\fill}$\Box$\medskip\medskip\par}

\newcommand{\n}{\rm}

\newcommand{\R}{{\mathbb R}}
\newcommand{\F}{{\mathbb F}}

\newcommand{\Q}{{\mathbb Q}}

\newcommand{\Z}{{\mathbb Z}}
\newcommand{\C}{{\mathbb C}}

\newcommand{\K}{{\mathbb K}}

\newcommand{\A}{{\mathbb A}}

\newcommand{\oo}{{\mathfrak o}}
\newcommand{\uu}{{\mathfrak u}}

\newcommand{\Gl}{\mathop{\rm Gl}\nolimits}

\newcommand{\Hom}{\mbox{\n Hom}}

\newcommand{\Sp}{\mbox{\n Sp}}

\newcommand{\Aherm}{{\mbox{\n Aherm}}}

\newcommand{\GL}{\mbox{\rm GL}}

\newcommand{\UU}{\mbox{\rm U}}

\newcommand{\M}{\mbox{\rm M}}

\newcommand{\pr}{\mbox{\rm pr}}

\newcommand{\XX}{{\cal X}}

\newcommand{\Herm}{{\rm Herm}}

\newcommand{\PP}{\Bbb{P}}
\newcommand{\AAA}{\Bbb{A}}
\newcommand{\HH}{{\cal{H}}}

\newcommand{\eps}{{\varepsilon}}
\newcommand{\OO}{{\rm O}}

\newcommand{\msk}{\medskip}
\newcommand{\ssk}{\smallskip}
\newcommand{\nin}{\noindent}

\begin{document}

\title{On the Hermitian Projective Line as a Home for the
Geometry of Quantum Theory}

\author{Wolfgang Bertram\footnote{\footnotesize
Institut \'Elie Cartan Nancy, Nancy-Universit\'e, CNRS, INRIA, 
Boulevard des Aiguillettes, B.P. 239, F-54506 Vand\oe{}uvre-l\`{e}s-Nancy,
 France;
{\tt bertram@iecn.u-nancy.fr}  } }

\maketitle

{\footnotesize
{\noindent{\bf Abstract.\/}}
In the paper ``Is there a Jordan geometry underlying quantum physics?''
 \cite{Be08}, {\em generalized projective geometries}
have been proposed as a framework for a geometric formulation of
Quantum Theory. In the present note, we refine this proposition
by discussing further structural features of Quantum Theory:
the link with {\em associative involutive algebras $\AAA$} and with 
{\em Jordan-Lie}
and {\em Lie-Jordan algebas}. The associated geometries are
{\em (Hermitian) projective lines over $\AAA$}; their axiomatic
definition and theory will be given in subsequent work with 
M.\ Kinyon \cite{BeKi08}. 
}

\msk

{\footnotesize 



\msk



\section*{Introduction}

\subsection{The geometry of quantum theory}

The quest for the ``Geometry of Quantum Theory'' goes back almost to
the days when John von Neumann laid the axiomatic foundations of the theory: 
it seems that von Neumann himself was not entirely satisfied by the
non-geometric and linear character of the axiomatic foundations
of Quantum Theory --  together with G.\ Birkhoff he tried to base
the theory on more fundamental and geometric concepts called the ``logic''
of quantum theory; the beautiful book ``The geometry of quantum theory''
 \cite{Va85} gives a full account on these and subsequent developments.
As a sort of conclusion, the author says (loc.\ cit., p.\ 6):
``...quantum mechanical systems are those whose logic form some sort of
projective geometry''. 

Coming from rather different starting points, various other authors have
arrived at  similar conclusions: they
%
propose to model the geometry of quantum theory
on infinite dimensional manifolds, among which
infinite dimensional projective spaces $\PP \HH$ (the space of pure
states, where $\HH$ is an infinite 
dimensional Hilbert space)  play
a central r\^ole, see, e.g.,  \cite{Ki79}, \cite{AS98}, \cite{BH01},
\cite{CGM03}. 
In  \cite{CGM03}, this approach, named ``delinearization program'', is 
motivated as follows:
``The delinearization program, by itself, is not related in our opinion
to attemps to construct a non-linear extension of QM with operators that
act non-linearly on the Hilbert space $\HH$. 
The true aim of the delinearization program is to free the mathematical
foundations of QM from any reference to linear structure and to linear
operators. It appears very gratifying to be aware of how naturally
geometric concepts describe the more relevant aspects of ordinary QM,
suggesting that the geometric approach could be very useful also in solving
open problems in Quantum Theories.''

In the paper \cite{Be08}, the present author
proposed an approach following the same general principles, 
but with the significant
difference that we try to geometrize rather the space of {\em observables}
as primary object, and not so much the space of {\em (pure) states}.
Remarkably, the resulting geometries still share many features with
the projective spaces $\PP  \HH$ ; we therefore call them {\em
generalized projective geometries}.
These form an interesting and quite large category that seems to be 
suitable for a geometrical formulation of some aspects of quantum theory.
However, this framework still is too general --
 as already pointed out in \cite{Be08}, it seems that
Nature has chosen among these geometries a fairly special one, namely
a geometry that resembles in many respects  a projective {\em line}. 
Of course, by this we do not mean
a usual projective line $\K \PP^1$ over a commutative
field $\K$, but rather a kind of  projective
line over an infinite dimensional $*$-algebra $\AAA$, 
called the {\em Hermitian projective line}. 
In this note, we will present their definition, as well as their mathematical
genes which come from the theory of associative and {\em Jordan} algebras.

\subsection{Jordan and Lie structures}

Our definition of generalized projective geometries has its origin in
{\em Jordan theory}:
the associative product $xy$ in an associative algebra can be
decomposed into a symmetric and a skew-symmetric part:
$$
xy = \frac{xy+yx}{2} + \frac{xy-yx}{2} =: x \bullet y + \frac{[x,y]}{2}.
\eqno(0.1)
$$
The skew-symmetric part $[x,y]$ gives rise to a {\em Lie algebra}, and
the symmetric part $x \bullet y$ to a {\em Jordan algebra}.
Axiomatically, since the foundational work of Pascual Jordan \cite{J32},
these algebras are defined by the following two properties (cf.\ \cite{McC04})

\ssk
(J1) $x \bullet y=y \bullet x$ (commutativity),

(J2) $x \bullet (x^2 \bullet y) = x^2 \bullet (x \bullet y)$ (the Jordan
identity).

\ssk \nin
Not every Jordan algebra is a subalgebra of an associative algebra with 
respect to the symmetrized product; if this is the case, the Jordan algebra
is called {\em special}. For instance, the Jordan algebra of observables in
Quantum Mechanics, the algebra $\Herm(\HH)$ of Hermitian operators, clearly
is special. 
In contrast to the case of Lie algebras, it is not easy to
get a feeling for the nature of Jordan algebras in an axiomatic approach
based on the defining identities (J1) and (J2).  
Indeed, in the author's opinion, it is much more
appropriate to start to study some more general objects whose structure
is, in a certain sense, much simpler, namely {\em Jordan triple systems}
and {\em (linear) Jordan pairs}.
They are easily interpreted in terms of {\em $3$-graded Lie algebras},
see \cite{Be08} or \cite{Be07} for an elementary exposition.

\ssk
Geometrically, Jordan pairs correspond to generalized projective geometries
in a similar way as Lie algebras correspond to Lie groups,
and Jordan triple systems correspond to such geometries together with
a suitable involution, called
{\em (generalized) polar geometries}.
Jordan {\em algebras} are fairly complicated objects since the 
corresponding geometries carry
all the preceding structures, plus an additional one, called an {\em
absolute null system}. 
This means, roughly, that the geometry is {\em canonically}
isomorphic to its dual geometry. For instance,
among ordinary projective spaces $\K \PP^n$, only
the projective line $\K \PP^1$ has this feature: it is {\em canonically}
isomorphic to its dual projective space of hyperplanes (since only in
dimension 1 a hyperplane is a point!). 
Therefore geometries corresponding to Jordan algebras can be considered
as ``non-associative generalizations of the projective line'' -- see
\cite{Be08} for all this. 

\ssk
Now, as already mentioned above, the Jordan algebra of Quantum Mechanics,
$\Herm(\HH)$, is {\em special}, and it even is very special in the sense
that it is a space of  {\em Hermitian elements in an associative $*$-algebra}.
Therefore there must be some additional
geometric structure on the corresponding ``projective line'', corresponding to
the additional algebraic structure given by the  {\em associative}
product: we may call it ``associative geometry''. What sort of
geometry is this?
Surprisingly, it seems that this question has never been seriously
investigated. The purpose of the present Note is to give the necessary
mathematical definitions and background for its understanding; the
``associative geometry'' itself will be axiomatically defined
and investigated in subsequent work with Michael Kinyon (\cite{BeKi08}).

\ssk
Algebraically, ``associative geometry'' in this sense
 is closely related to concepts
of ``Jordan-Lie'' and ``Lie-Jordan'' algebras that have appeared in
the literature, and which in a sense
are axiomatic versions of decompositions, like (0.1), of the associative product. 
We recall the basic definitions and give some comments on them (Chapter 2); 
a very remarkable feature is that they introduce a ``coupling constant''
$C$ measuring the way in which the Jordan- and Lie-structures are
linked to each other. For $C=0$, we essentially get commutative 
Poisson-algebras, whereas for $C=1$ and $C=-1$ we get two different 
properly ``quantum'' structures. 
The ``classical limit'' $C \to 0$ thus is interpreted as 
``going from commutative non-associative to commutative associative''.
This calls for comparison with  the philosophy of 
Non-commutative Geometry -- see the final Chapter 3 for some concluding
remarks.

\bigskip \nin
{\bf Acknowledgments.}
Parts of the present work have been presented at the 
XXVII Workshop on Geometrical Methods in Physics in Bia\l owie\.za 2008.
I would like to thank the organizers for their kind invitation and
great hospitality.

\section{Geometry of the Hermitian projective line}

\subsection{The  projective line over a ring}

Let  $\AAA$ be an associative
algebra, defined over some commutative base field or ring $\K$.
The {\em projective line
over $\AAA$} is, by definition (cf., e.g., the article by A.\ Herzer in
\cite{Bue95}), the set
$\AAA\PP^1$ of all
submodules $x$ of the right $\AAA$-module 
$\AAA \oplus \AAA$ that are isomorphic to $\AAA$
and admit a complementary submodule $x'$ isomorphic to $\AAA$.
The projective line $\AAA \PP^1$ is non-empty since it contains at least
the two elements
$$
o^+:= 0 \oplus \AAA = e_2 \AAA \, \mbox{(second factor)}, \quad
o^-:= \AAA \oplus 0 = e_1 \AAA  \, \mbox{(first factor)}.
$$
The general $\A$-linear group 
$G:=\GL(2,\AAA)$
acts in the usual way on $\AAA^2$ from the left; this action permutes
all $\AAA$-right modules and defines a transitive action of $G$ on
$\AAA \PP^1$: if $x \in \AAA \PP^1$ with base vector $v$, having a 
complement $x'$ with base vector $v'$, just let $g$ be the $\AAA$-linear
map sending $e_2$ to $v$ and $e_1$ to $v'$; then $g o^+ = x$ (and
$go^- = x'$).
The stabilizer of $o^+$, resp.\ of $o^-$, is the subgroup
$P^-$ of lower (resp.\ upper) triangular matrices in $G$, so that
$\AAA \PP^1$ can be written as a homogeneous space
$$ 
\XX^+ := G.o^+ = G/P^-, \quad \XX^- := G.o^- = G/P^+.
$$
Of course, $\XX^+ = \XX^- = \AAA \PP^1$
 as sets; but the base points are different.

We say that a pair $(x,y) \in \AAA \PP^1 \times \AAA \PP^1$
 is {\em transversal} (in incidence geometry one also uses the term
 {\em remote}, cf.\ \cite{Bue95}, loc.\ cit., meaning ``as non-incident as
possible''), and
then write $x \top y$, if $x$ and $y$ are complementary subspaces:
$$
\AAA \oplus \AAA = x \oplus y 
$$
%
We sometimes write $(\XX^+ \times \XX^-)^\top$ 
for the set of transversal pairs.
This set is non-empty since the pair $(o^+,o^-)$ is transversal, and
 the action of $\Gl(2,\AAA)$ on it  is 
 transitive (just define the matrix $g$ as above).
The stabilizer of the canonical base point $(o^+,o^-)$ is
the subgroup $H$ of $G$ consisting of diagonal matrices:
$$
(\XX^+ \times \XX^-)^\top = G.(o^+,o^-) = G/H.
$$

The projective line can be seen, in a natural way, as a ``projective
completion of the algebra $\AAA$'':
let, for any $y \in \AAA \PP^1$,
$$
y^\top := \{ x \in \AAA \PP^1 | \, x \top y \},
$$
the set of elements that are transversal to $y$.
Then $y^\top$ is, in a natural way, an affine space over $\K$ 
isomorphic to $\AAA$: 
indeed, since the action of $\Gl(2,\AAA)$ is transitive, without loss
of generality we may assume that $y=o^-$ is the first factor. 
But all complements of the first factor have a unique base vector of
the form $(a,1)$ with $a \in \AAA$. In other words,
the subgroup of matrices
$$
\begin{pmatrix}
1 & a \cr 0 & 1 \cr
\end{pmatrix}, \quad a \in A,
$$
acts simply transitively on the set of complements of the first factor.
In particular, we may canonically identify
$\AAA$ with $(o^+)^\top$ or with $(o^-)^\top$, and regard the projective line
as some sort of ``projective completion'' of $\AAA$.

\msk 
\nin {\sl Examples (cf.\ \cite{Be07}, \cite{Be08}, \cite{BeNe05}).}

\nin {\bf (1)}
 Assume  $\AAA$ is a skew-field, or, in other words, an
associative division algebra over $\K$. Then we can write
$\AAA \PP^1 = \AAA \cup \{ \infty \}$, where $\infty = o^-$ is
the ``unique point at infinity of the affine part $\AAA$''.
In particular, if $\AAA = \R,\C$ or another locally compact
topological field, then $\AAA \PP^1$ is the one-point compactification
of $\AAA$.

If $\AAA$ is not a skew-field, the ``set at infinity'' has more than
just one element, and its geometric structure is richer and more interesting. 

\ssk
\nin {\bf (2)} 
If $\AAA = \R[\eps]$, $\eps^2=0$, is the ring of dual numbers
over $\R$, then $\AAA \PP^1$ is the tangent bundle $T(\R \PP^1)$ of
the usual real projective line. The set at infinity is the tangent space
$T_\infty (\R \PP^1)$ at the point $\infty$ of $\R\PP^1$.

\ssk
\nin {\bf (3)}
If $\AAA = M(n,n;\K)$ is the matrix algebra over a commutative
unital ring $\K$, then $\AAA \PP^1$ is naturally isomorphic to the
Grassmannian manifold of  $n$-spaces in $\K^{2n}$.

\ssk
\nin {\bf (4)}
 If $\AAA= F(M,\K)$ is the commutative algebra of all functions
from a set $M$ to the commutative base ring $\K$, then $\AAA \PP^1$
is the space of all functions from $M$ to the projective line
$\K \PP^1$. 

\ssk
\nin {\bf (5)}
If $\AAA$ is a  principal ideal ring,  then
$\AAA \PP^1$ is the one-point completion of the quotient field $\F_\AAA$ of
$\AAA$; in other words, it is the projective line over the field
$\F_\AAA$. Indeed, let $q \in \F_\AAA$. 
If $q=0$, we associate to it the point $x=[\pr_1]=o^-$ of $\AAA \PP^1$. Else write
$q=\frac{s}{r}$  with $r$ and $s$ relatively prime in $\AAA$, 
so there exist $a,b \in \AAA$
with $ar-bs=1$, hence the matrix
$g:=\begin{pmatrix} s & a \cr r & b \cr \end{pmatrix}$
is invertible. The  point $x:= g.[\pr_1]
\in \AAA \PP^1$ (submodule with base vector $(s,r)$) only depends on $q$.
Conversely, let $x \in \AAA \PP^1$ generated by $(s,r) \in \AAA^2$.
Then the vector $(s,r)$ can be completed by a vector
 $(a,b) \in \AAA^2$ to a matrix $g$ as above with determinant equal to $1$.
If $s=0$, then $x=[\pr_1]$; if $r=0$, then $x=[\pr_2]$, and in all other
cases $x$ can be identified with the element $q=\frac{s}{r}$ 
of the quotient field. Both constructions are inverse to each other, and thus
we have a bijection between $\F_\AAA \cup \{ \infty \}$ and $\AAA \PP^1$.

For instance, $\Z \PP^1$ is the rational projective line $\Q \PP^1$, and
 if $\AAA=\K[X]$ is the polynomial ring over a field $\K$, then
$\AAA \PP^1 = \K(X) \cup \{ \infty \}$ is the completion of the rational
function field by a ``function'' $\infty$ which can be considered as the
inverse of the zero function. In both cases, the ``set at infinity''
is very big: it is a sort of infinite dimensional manifold over $\AAA$.

\ssk
\nin {\bf (6)}
Let us assume that $\AAA$ is a {\em continuous inverse algebra (c.i.a.)}:
a topological algebra over a topological ring
$\K$ such that the unit group $\AAA^\times$ is open in $\AAA$ and inversion
$i:\AAA^\times \to \AAA$ is a continuous map. We assume also that
$\K^\times$ is dense in $\K$ (in particular, the topology is not discrete).
Then inversion is actually smooth over $\K$, and
 $\AAA \PP^1$ is a smooth manifold over $\K$, modelled on the topological
linear space $\AAA$ (see \cite{BeNe05}). 
For instance, $\AAA$ may be any Banach algebra over
$\K=\R$, $\C$ or $\Q_p$.

\subsection{The Hermitian projective line}

Next we consider an associative algebra $\AAA$ with an {\em involution} 
 $*:\AAA \to \AAA$, $a \mapsto a^*$ 
(antiautomorphism of order $2$ stabilizing $\K=\K 1$;
following standard terminology \cite{Le06} we say that $*$ is {\em
of the first kind} if $*$ induces the identity on $\K$, and
{\it of the second kind} else).  Then the involution $*$
 lifts to an
involution of the projective line $\AAA \PP^1$ whose fixed point set
is called the {\em Hermitian projective line}, see \cite{BeNe05}. 
Let us give here a slightly modified version of the construction
given in loc.\ cit.:
for any matrix $B=(b_{ij}) \in M(2,2;\AAA)$ we may define a sesquilinear
form $\beta=\beta_B$ on $\AAA^2$ by
$$
\beta(x,y) = \sum_{i,j=1}^2 x_i^* b_{ij} y_j.
$$
The sesquilinearity property reads
$$
\forall \lambda,\mu \in \AAA : \quad \quad
\beta\bigl( x \lambda ,y \mu \bigr) = \lambda^* \beta(x,y) \mu,
$$
hence the orthogonal complement $E^{\perp,\beta}$
 of an $\AAA$-right submodule
$E$ is again an $\AAA$-right submodule.
Assume now that $\beta$ is {\em non-degenerate}, i.e., $B$ is an invertible
matrix. Then,  if 
$E \in \AAA \PP^1$,  also $E^{\perp,\beta} \in \AAA \PP^1$: indeed,
if $\AAA \oplus \AAA = x \oplus y$ is a direct sum decomposition
with both factors $x$ and $y$ isomorphic to $\AAA$, then so
is $\AAA \oplus \AAA = x^\perp \oplus y^\perp$.
Moreover, the map thus defined
$$
\perp^\beta: \AAA \PP^1 \to \AAA \PP^1, \quad E \mapsto E^{\perp,\beta}.
$$
is a bijection (with inverse corresponding to $B^{-1}$). 
It is $G$-equivariant in the following sense:
$(g.x)^\perp = \phi(g)^{-1}.x^\perp$ where 
$\phi$ is the anti-automorphism ``$\beta$-adjoint'' of $\M(2,2;\AAA)$ given by
$$
\phi \begin{pmatrix} a & b \\ c & d \\ \end{pmatrix} =
B^{-1} 
\begin{pmatrix} a^* & c^* \\ b^* & d^* \\ \end{pmatrix} B .
$$
Now let us consider the following
three sesquilinear forms on $\AAA^2$ given by
$$
\begin{matrix}
\omega\bigl( (x_1,x_2),(y_1,y_2) \bigr) & = & x_1^* y_2 - x_2^* y_1, \cr
\vartheta\bigl( (x_1,x_2),(y_1,y_2) \bigr) & = & x_1^* y_2 + x_2^* y_1, \cr
\sigma\bigl( (x_1,x_2),(y_1,y_2) \bigr) & = & x_1^* y_1 - x_2^* y_2, \cr
\end{matrix}
$$
corresponding to the three matrices
$$
\Omega:= \begin{pmatrix} 0 & 1 \\ -1 & 0 \\ \end{pmatrix}, \quad
T:= \begin{pmatrix} 0 & 1 \\ 1 & 0 \\ \end{pmatrix}, \quad
S:= I_{1,1}=  \begin{pmatrix} 1 & 0 \\ 0 & -1 \\ \end{pmatrix},
$$
called the ``$*$-symplectic'', ``$*$-hyperbolic'', ad ``$*$-symmetric'' forms.
Since these forms are Hermitian, resp.\ skew-Hermitian, the corresponding
maps $\perp^\beta$ are involutions, i.e., of order $2$.
The fixed point sets
$$
\begin{matrix}
\PP_h & := & (\AAA \PP^1)^{\perp,\omega} = \{ E \in \AAA \PP^1 | \,
E^{\perp,\omega}=E \}, \cr
\PP_{sh} & :=& (\AAA \PP^1)^{\perp,\vartheta} = \{ E \in \AAA \PP^1 | \,
E^{\perp,\vartheta}=E \}, \cr
\PP_{u} & := & (\AAA \PP^1)^{\perp,\sigma} = \{ E \in \AAA \PP^1 | \,
E^{\perp,\sigma}=E \} \cr
\end{matrix}
$$
are  called the {\em Hermitian}, respectively {\em
skew-Hermitian} and {\em unitary projective line over the involutive
algebra $(A,*)$}. 
To justify the last terminology, note that the {\em $*$-unitary group}
$$
\UU(\AAA,*):=\{ a \in \AAA | \, a^* a = a a^* = 1 \}
$$
is imbedded into $\PP_u$ via $a \mapsto (1,a) \AAA$
(to see this, just note $\sigma\bigl((1,a),(1,a)\bigr)=1 - a^* a =0$).
In some cases, this imbedding is a bijection (see examples below).
Of course, via a base change the forms $\vartheta$ and $\sigma$ are isomorphic,
and therefore also the skew-Hermitian and the unitary projective line are
isomorphic. In general, they are not isomorphic to the Hermitian projective
line; but in some interesting special cases they are (see below).

\ssk
Let us denote by $\XX$ one of the three kinds of projective line
$\PP_h$, $\PP_{sh}$, $\PP_u$ just defined.
There are some general features already encountered for the full projective
line $\AAA \PP^1$ that carry over: the notion of {\em transversality in $\XX$} 
remains the same;
we have a transversal pair of base points: for $\PP_h$ and
$\PP_{sh}$ it is again $(o^+,o^-)$; for $\PP_u$ we rather have to take the
two diagonals in $\AAA \oplus \AAA$. There is a natural group $G$ acting, 
namely
the  unitary groups of the respective forms.
For $\omega$, we call it the {\em $*$-symplectic group},
denoted by $\Sp(\AAA,\omega)$; for $\vartheta$, the {\em $*$-pseudo unitary
group}, denoted by $\UU(\AAA,\AAA,*)$. One can show that, for $\omega$,
this action is always transitive both on the Hermitian projective line and on
the set of transversal pairs, so we may consider homogeneous spaces
of the form $G/P^-$, $G/P^+$, $G/H$ as before,
whereas for $\vartheta$, this need not always be the case: in this case
we better write 
$$
\XX^+:= \UU(\AAA,\AAA,*).o^+, \quad \XX^-:= \UU(\AAA,\AAA,*).o^-,
$$
and these two orbits may be equal or disjoint in the skew-Hermitian projective
line. In any case, it remains true that the set $y^\top$ of all transveral
elements in $\XX$ to a given element $y$ is always an affine space over $\K$,
which now is modelled on the sets
$$
\Herm(\AAA,*):= \{ a \in \AAA | \, a^* = a \}, \quad \mbox{respectively}
\quad \Aherm(\AAA,*):= \{  a \in \AAA | \, a^* = - a \}
$$
for the Hermitian, resp.\ skew-Hermitian, projective line.

\msk \nin
{\sl Examples (cf.\ \cite{Be07}, \cite{Be08}, \cite{BeNe05}).}

\ssk
\nin {\bf (1)} If $\AAA$ is commutative, then $\Herm(\AAA,*)$ is also
a commutative algebra, and $\PP_h$ is just the projective line over
this algebra. If $\AAA = F(M,\K)$ is a function algebra, then
$\PP_u$ is the group of functions with values in the ``circle group''
$\{ r \in \K^\times | \, r^* = r^{-1} \}$.

\ssk \nin {\bf (2)}
If $\AAA=M(n,n;\K)$ and $X^*=X^t$, then $\PP_h$ is the variety of
Lagrangian subspaces of the canonical symplectic form on $\K^{2n}$,
and $\PP_{sh}$ is the Lagrangian variety for the quadratic form of
signature $(n,n)$. For $\K=\R$, the imbedding of the orthogonal group
$\OO(n)$ into the latter is a bijection.


\subsection{The unitary-Hermitian projective line}

Quantum mechanics requires to work over the field $\C$
of complex numbers and the involution $*$ to be $\C$-antilinear.
This has the particular consequence that the spaces of
Hermitian elements ($a^*=a$) and skew-Hermitian elements ($a^*=-a$)
are isomorphic. In general, let us call an involution $*$ of an associative 
$\K$-algebra $\AAA$ {\em of complex type} if there exists an element
$i \in \K$ with $i^2 = -1$ and $i^* = -i$.
Then multiplication by $i$ is a $\K$-linear isomorphism from
$\Herm(\AAA,*)$ onto $\Aherm(\AAA,*)$.
The diagonal matrix $\mbox{dia}(i,1) \in \Gl(2,\AAA)$
then induces a bijection from the Hermitian projective line onto
the skew-Hermitian one.
Mathematically, this is a very
special feature; but nevertheless 
Nature has chosen it as being part of the structure of
Quantum Mechanics:

\msk \nin
{\sl Example.}
Let $\AAA = M(n,n;\C)$ and $X^*=\overline X^t$. Then $\PP_h$, $\PP_{sh}$
and $\PP_u$ are all isomorphic to the variety of all Lagrangian subspaces of 
the Hermitian form on $\C^{2n}$ with signature $(n,n)$. The imbedding of
the unitary group $\UU(\AAA;*)=\UU(n)$ into $\PP_u$ is in this case
a bijection. The ``big'' group $G$ acting on $\PP_h \cong \UU(n)$ 
is $\PP \UU(n,n)$.
To get the setting of Quantum Mechanics, one may in this example
replace $M(n,n;\C)$ by the bounded operators
on an infinite dimensional Hilbert space $\HH$ (unbounded operators
can be dealt with by a more subtle choice of {\em Jordan pair}, see below).

One may also  replace the positive involution considered above by the
``indefinite involution''
$X^*:= I_{p,q} \overline X^t I_{p,q}$, where 
$I_{p,q}$ is the usual diagonal matrix of signature $(p,q)$ and square one.
As spaces, $\PP_h$, $\PP_{sh}$ and $\PP_u$ are then still  the same as above,
the only thing that changes is that we now consider another
``polarity'', which corresponds to the imbedding of the pseudo-unitary group 
$\UU(\AA,*)=\UU(p,q)$ as open dense set into the compact space
$\PP_u$ (in the Russian literature this is
 called the ``Potapov-Ginzburg transformation'').

\subsection{Positivity}

There is another particular feature of Quantum Mechanics that one
has to take account of: {\em positivity}.
Algebraically, this corresponds to the positivity condition in the
definition of a $C^*$-algebra ($|| xx^*|| = ||x||^2$);
geometrically, it corresponds to the fact that there exists a partial
order on the Jordan algebra $\Herm(\AAA,*)$ 
such that squares are positive (in particular,
$\Herm(\K,*)$ then has to be an ordered ring or field), and also to
the fact that (under some additional conditions)
 the imbedding of the unitary group
$\UU(\AAA,*)$ into $\PP_u$ becomes a bijection.
These are interesting topics for further work; however,
for the logical development of the theory it seems useful not to
introduce such positivity assumptions at an early stage.

\section{Jordan-, Lie- and Jordan-Lie algebras}

\subsection{Jordan pairs and triple systems}

For the sake of completeness, we just recall here the mere definitions:
a {\em (linear) Jordan pair} is a pair $(V^+,V^-)$
of $\K$-modules together with two trilinear maps
$T^\pm : V^\pm \times V^\mp \times V^\pm \to V^\pm$ satisfying the
following identities (LJP1)
and (LJP2):

\begin{description}
\item{(LJP1)} $T^\pm(x,y,z) = T^\pm(z,y,x)$,
\item{(LJP2)} $T^\pm(a,b,T^\pm(x,y,z)) =$
\item{ } $\quad \quad \quad \quad \quad T^\pm(T^\pm(a,b,x),y,z) -
T^\pm(x,T^\mp(b,a,y),z) + T^\pm(x,y,T^\pm(a,b,z))$.
\end{description}

\nin The basic example of a linear Jordan pair is given by spaces of
rectangular matrices: 
$$
(V^+,V^-) = \bigl(\Hom(F,E),\Hom(E,F)\bigr) \quad \mbox{with} \quad
T^\pm(x,y,z)=xyz+zyx.
$$
By definition, a {\em Jordan triple system} (JTS) is a $\K$-module $V$ together
with a trilinear map $T:V \times V \times V \to V$ satisfying the
identities
(JT1) and (JT2) obtained from (LJP1) and (LJP2) by omitting the
superscripts $\pm$. The basic example is again the space of rectangular matrices,
$$
V=M(p,q;\K) \quad \mbox{with} \quad
T(x,y,z)=xy^t z + zy^t x.
\eqno (2.1)
$$
Note that the transpose corresponds to the choice of a scalar product.
If $p=q$, or if $V=\AAA$ is any associative algebra, one also has another
Jordan triple product given by $T(a,b,c)=abc+cba$. 
In general, every JTS gives rise to a Jordan pair $(V^+,V^-):=(V,V)$
with $T^+=T^-=T$ (but not every Jordan pair is of this form),
and every Jordan algebra (with product $\bullet$) gives rise to a JTS via
$$
T(x,y,z) = \frac{1}{2}\bigl(x \bullet (y \bullet z) - y \bullet (x \bullet z)
+ (x \bullet y) \bullet z \bigr) \,  .
\eqno (2.2)
$$
Summing up, there are several functors between the following
categories:  associative algebras;
Jordan algebras; Jordan triple systems; Jordan pairs (and several others,
such as: Lie triple systems; associative and alternative pairs, etc). 
Each of these functors sheds light on certain features of the categories
between which it is defined.


\subsection{Jordan-Lie algebras}

As already noticed above, some algebras carry simultaneously
the structure of a Jordan and of a Lie algebra: on the one hand, we have
the full associative algebras $\AAA$ with the usual (anti-) commutators;
on the other hand, the spaces $\Herm(\HH)$ of Hermitian operators in
a complex Hilbert space $\HH$, where a factor $i$ comes in. 
The concept of a {\em Jordan-Lie algebra} takes account of both
cases; the definition is due to G.\ Emch (\cite{E84}), although
 the concept seems to have appeared first in the paper
\cite{GP76}.  

\msk \nin
{\bf Definition.}
Let $C \in \K$ be a constant.
A $\K$-module $V$, equipped with two bilinear products $[x,y]$ and
 $x \bullet y$ is called a {\em Jordan-Lie algebra
(with coupling constant $C$)}
if

\begin{description}
\item{(JL1)} $(V,[ \cdot,\cdot ])$ is a Lie algebra;
\item{(JL2)} $(V,\bullet)$ is a Jordan algebra;
\item{(JL3)} the Lie algebra acts by derivations of the Jordan algebra, 
that is,
$$
[x,u \bullet v] = [x,u] \bullet v + u \bullet [x,v] ,
$$
\item{(JL4)} the associators of both products are proportional:
$$
(x \bullet y) \bullet z - x \bullet (y \bullet z) = -C \bigl(
[[x,y],z]-[x,[y,z]] \bigr).
$$
\end{description}

\nin
Of course, (JL4) can also be written
$
(x \bullet y) \bullet z - x \bullet (y \bullet z) = C [[z,x],y],
$
thanks to the Jacobi identity.
The main examples are:

\msk \nin
{\bf (1)} Commutative Poisson algebras: for $C=0$, Condition (JL4)
says that $\bullet$ is a commutative and associative product,
on which the Lie algebra acts by derivations, by (JL3).

\ssk \nin
{\bf (2)} Associative algebras $V=\AAA$ with usual Jordan product
 and Lie bracket:
(JL3) is clear since the Lie algebra already derives the associative product;
(JL4) with $C=4$ follows by a direct calculation. 

\ssk \nin  
{\bf (3)} Hermitian elements:
under the assumptions from Section 1.3, let
$V=\Herm(\AAA,*)$ with its usual Jordan product and the modified Lie
bracket $[x,y]:=i (xy-yx)$. The same calculation as in the preceding
example yields an additional factor $i^2=-1$ on the right hand side of
(JL4), whence we get a Jordan-Lie algebra with coupling constant $C=-4$.

\msk 
Conversely, 
given a Jordan-Lie algebra with coupling constant $C$, consider
the scalar extension $R:=\K[X]/(X^2 - C)$ of $\K$; writing $i:=[X]$,
 this is simply
the ring $\K \oplus i \K$ with defining relation $i^2 =C$.
 (For $C=0$, these are the dual numbers over $\K$.)
Let $V_R:=V \oplus iV$ the scalar extension of the given Jordan-Lie
algebra, which is again a Jordan-Lie algebra, now defined over $R$.
Define a new product on $V_R$ by
$$
xy:= x \bullet y + i [xy].
$$
By a direct calculation (cf.\ \cite{E84}, p.\ 307) one sees that the
associator of this product is
$$
(xy)z-x(yz)=C [[x,z],y] - i^2 \bigl( [[x,y],z] - [[y,z],x] \bigr) = 
(C- i^2)[[x,z],y] =  0 ,
$$
hence $V_R$ is an associative algebra. The ``conjugation map''
$a+ib \mapsto a-ib$ is an involution of this algebra.
If $C$ is a square in $\K^\times$, we get back  Example (2),
and if $-C$ is a square in $\K^\times$, we get back  Exemple (3).
If $C=0$, we have constructed an associative, in general non-commutative
algebra out of a commutative Poisson algebra; it is a sort of first
approximation of a deformation quantization of that algebra. 

\ssk
Categorial notions for Jordan-Lie algebras follow the usual pattern:
homomorphisms, ideals, simple, semi-simple objects are those which
have the corresponding properties both for the Jordan and the Lie
product; for invertible $C$, simple objects then correspond
to simple associative algebras over $R$ with involution.
For $\K=\R$, the classification of simple finite-dimensional objects
 is therefore very easy: for $C>0$, we get simple associative algebras
(that is, matrix algebras over the three associative
 real division algebras, by the
classical Burnside theorem); for $C<0$, we have to look at simple
complex algebras $M(n,n;\C)$ with $\C$-antilinear involution: it is known that
all such involutions correspond to adjoints with respect to a non-degenerate
Hermitian form on $\C^n$ (see \cite{Le06}). Therefore the Jordan part
of a simple finite-dimensional real Jordan-Lie algebra is isomorphic
to $V = \Herm(p,q;\C)$ and its Lie part isomorphic to $\uu(p,q)$,
the Lie algebra of the pseudo-unitary group $\UU(p,q)$.
In infinite dimension, the classification of Jordan-Lie algebras
contains the classification of $C^*$-algebras as a subproblem.

\ssk
We add two remarks on important special features of Jordan-Lie algebras:

\ssk \nin
{\bf 1.} Jordan-Lie algebras permit to single out {\em complex} Hermitian matrices
by a {\em purely real} concept. This means that an axiomatic approach
to quantum mechanics without making use of complex numbers is possible.
See also \cite{La93} and \cite{La94}.

\ssk \nin
{\bf 2.} {\em Tensor products} exist in the category of Jordan-Lie algebras.
This is very remarkable since tensor products  neither exist in the
category of Jordan algebras nor in the one of Lie algebras. 
This observation is the starting point of the paper \cite{GP76},
where the notion of {\em composition class} as a class of two-product
algebras closed under tensor products is introduced;  the
idea to characterize quantum and classical mechanics as certain
composition classes  goes back to Niels Bohr.

\subsection{Lie-Jordan algebras}

The folllowing definition is (for $C=1$)
 due to Grishkov and Shestakov \cite{GS01}:
Let $C \in \K$ be a constant.
A $\K$-module $V$, equipped with a bilinear product $[x,y]$ and a
trilinear product $T:V^3 \to V$  is called a {\em Lie-Jordan algebra
(with coupling constant $C$)}
if

\begin{description}
\item{(LJ1)} $(V,[ \cdot,\cdot ])$ is a Lie algebra,
\item{(LJ2)} $(V,T)$ is a JTS,
\item{(LJ3)} the Lie algebra acts by derivations of the JTS
$T$, that is,
$$
[x,T( u,v,w)] = T( [x,u], v,w ) + T (u, [x,v],w ) + T( u,v,[x,w] ),
$$
\item{(LJ4)} skew-symmetrized $T$ is proportional to the triple Lie bracket:
$$
T(x,y,z) - T(y,x,z)  = C [[x,y],z].
$$
\end{description}

\nin
Similar comments as in the preceding section can be made. 
Every Jordan-Lie algebra gives rise to a Lie-Jordan algebra, but the converse
is false: the $-1$-eigenspace of any involution of an associative algebra
$(\AAA,*)$ gives rise to a Lie-Jordan algebra (and 
Grishkov and Shestakov show
that every Lie-Jordan algebra is obtained in this way, if $C=1$).
For instance, if $\AAA = M(n,n;\R)$ and $X^* = X^t$, the $-1$-eigenspace
is the Lie algebra $\oo(n)$ of the orthogonal group, which is stable under
the usual Jordan triple product (2.1), giving rise to Lie-Jordan algebra; but it is not
a Jordan-Lie algebra since (2.1) never is obtained from a Jordan algebra via (2.2).

\section{Comments and afterthoughts}

{\bf 3.1.}
Jordan-Lie algebras (with a certain positivity condition, see \cite{E84})
are equivalent to $C^*$-algebras. Therefore
there is no contradiction between associative or non-associative
geometry in our sense and the philosophy of Non-commutative Geometry:
it is just a difference of language. Jordan geometry leads back to
a classical language for describing non-classical results; we re-introduce
the language of point-spaces on a level where Non-commutative Geometry 
teaches us to abandon point-spaces.
However, the point of view of Jordan-Lie algebras separates the Jordan and Lie
aspects of a $C^*$-algebra, and thus sheds light onto aspects that remain
unnoticed in a purely associative theory.

\msk \nin
{\bf 3.2.}
We think that Bohr's idea of ``composition classes'' (see Remark 2 in
Section 2.2) is interesting and deserves a careful re-investigation. 
Composition classes with $C=0$ are called ``classical'' and those with
$C$ invertible are called ``quantal''. 
Classical composition classes are thus commutative Poisson algebras;
quantal composition classes are generally non-associative, but also contain
associative commutative Jordan algebras (which, however, are then just
considered as commutative and not
 as Poisson algebras: if the left-hand side
of (JL4) is zero and $C$ is invertible, it follows that $[[V,V],V]=0$).
In \cite{GP76} (Appendix A)
also the case of symmetric and exterior tensor powers
is considered; only the symmetric ones give rise to a composition class.
Maybe the extension of such concepts to ternary products (Lie-Jordan algebras)
permits to include exterior powers in this picture.

\msk \nin
{\bf 3.3.}
In his work \cite{Pen89}, \cite{Pen05}, Roger Penrose frequently 
expresses the opinion that foundational problems of unifying
quantum theory with general relativity are related to the
coexistence of two different modes of time evolution in quantum
mechanics, the ``unitary Schr\"odinger evolution'' $\bf U$ and the
``state reduction'' $\bf R$. Mathematically, as already noticed
by G.\ Emch \cite{E84}, the ``Jordan part'' of an observable
somehow reflects its ``quantum aspect'', that is, $\bf R$, whereas
its ``Lie part'' rather represents $\bf U$.
Since  the concept of a Jordan-Lie algebra clearly separates these two
aspects, it seems to meet the
heart of the problem as formulated by Penrose.
This indicates that the ``coupling axiom'' (JL4) is of central importance
and should be better understood.
In particular, it is not at all easy to find a geometric interpretation of
this axiom on the level of the geometries of the corresponding projective
lines.

\vfill\eject
\end{document}